\documentclass[12pt,a4paper]{article}
\usepackage{fullpage}
\usepackage{authblk}
\usepackage{moreverb}
\usepackage{amsmath}
\usepackage{mathrsfs}
\usepackage{amsfonts}
\usepackage{amssymb}
\usepackage{bm}
\usepackage{enumerate}
\usepackage{tikz}
\usepackage{graphicx, multirow, subfigure, txfonts}
\usepackage{stfloats}

\usepackage[colorlinks,
            linkcolor=red,
            anchorcolor=blue,
            citecolor=green
            ]{hyperref}

\def\Z{\bm{ Z}}

\def\Z{{\mathbb{Z}}}

\numberwithin{equation}{section}
\newcommand{\argmin}{\mathop{{\rm arg}\min}}

\title{Robust Signal Reconstruction Using the Prolate Spherical Wave Functions and Maximum Correntropy Criterion \footnotemark[2]}

\author{Cuiming~Zou\thanks{zoucuiming2006@163.com}  }

\author{Kit Ian Kou\thanks{Corresponding author: kikou@umac.mo}}

\affil{\normalsize{Department of Mathematics, Faculty of Science and Technology, University of Macau, Macao, China}}
\date{}

\begin{document}
   \maketitle

\begin{abstract}
\normalsize

Signal Reconstruction is one of the most important problem in signal processing.
This paper proposes a novel signal reconstruction method based on the
prolate spherical wave functions (PSWFs) and maximum correntropy criterion (MCC).
The PSWFs are a kind of special functions, which have been proved having good performance in signal reconstruction.
However, the existing PSWFs based reconstruction methods only
consider the mean square error (MSE) criterion as the cost functions.
The MSE criterion is sensitive to the non-Gaussian noise, since it is builded up by the Gaussian assumption.
Therefore, for the impulsive noise or outliers, the MSE based reconstruction methods will lead to the large reconstruction error.
The proposed MCC and PSWFs based robust signal reconstruction method
 can reduce the impact of large and non-Gaussian noise.
The experimental results on the synthetic signals show that the proposed method can improve the MSE with notable gains in most cases.
\end{abstract}


\section{Introduction}
\label{Introduction}
Prolate spheroidal wave functions (PSWFs) are important functions in information and communication theory \cite{JJ2011}.
They, which is a special case of the spheroidal wave functions, possess many interesting  properties, such as double orthogonality in both the finite time domain and the whole real axis. The PSWFs are the most energy concentrated signals in energy concentration problem which was studied by Slepian \emph{et al}. \cite{SP1961, S1964, S1978}. In practical, their discrete forms also satisfy the orthogonality relations. The energy concentration problem aims to find the bandlimited functions with the maximum energy in a fixed time interval, which satieties the extreme conditions in the uncertainly principle \cite{K2004}.
The PSWFs are proved to be an orthogonal basis in the Paley-Wiener space \cite{JJ2011, MC2004}, which has extensively used for a variety of physical and engineering applications.

Most notably,  the PSWFs have been used successfully in sampling theory and signal reconstruction.
The famous Shannon's sampling theorem was created in 1949 \cite{S19491998},
which is the foundation of information theory. The reconstruction formula is
$f(x)=\sum_{k \in \Z} f(k W) \mbox{Sinc} \left( {x \over W}-k \right)$,
 which is known as the cardinal series expansion (basis functions obtained by appropriate
shifting and rescaling of the sinc-functions). Nowadays this theorem still plays a central role in signal, image processing and communication.
In \cite{WS2003,KG2003}, researchers studied some Shannon's reconstruction formulas associated with PSWFs.
The sinc-function $\mbox{Sinc}$ was introduced by the expansions of PSWFs' (also namely Slepian series) \cite{MZ2014, DK2016}.

In 2009,  Senay \emph{et al}. \cite{SCD2009} first utilized the PSWFs to the signal reconstruction problem.
They later extended the reconstruction method combining with the Tikhonov regularization in \cite{SOC2012}.
However, most existing signal reconstruction methods exploit
the mean square error (MSE) criterion as the cost functions due to the ease of analysis.
It is well known that the MSE is build by the hypothesis that the noise follows the Gaussian distribution.
In reality, the noises are more complicated and do not necessarily obey the Gaussianity assumption,
for example the impulsive noise \cite{PV2012}.
Once the assumption violates, the performance of the MSE based reconstruction methods
may severely decline.
In this paper, we propose a novel signal reconstruction method which is based on the
maximum correntropy criterion (MCC) in the information-theoretic learning \cite{LPC2007,P2010}.
Unlike the MSE, the MCC is independent of the noise distribution.
This makes our method more attractive in handling both Gaussian and non-Gaussian noise cases.

The paper is organized as follows. Section \ref{Preliminaries}
introduces some basic facts about  PSWFs and maximum correntropy criterion.
Section \ref{Signal Reconstruction} recalls the classical sampling theorem and relationship of sinc-functions and PSWFs.
We discuss the existing PSWFs based reconstruction methods and our proposed methods.
Section \ref{Experiments} presents the experimental results for uniformly sampling signal and non-uniformly
sampling signal. For both of the experiments, our proposed methods show good performances compare to the other related methods.
Some conclusions are drawn, and future works are proposed in Section \ref{sec.Conclusions}.

%%%%%%%%%%%%%%%%%%%%%%%%%%%%%%%%%%%%%%%%%%%%%%%%%%%%%%%%%%%%%%%%%%%%%%%%%%%%%%%%
\section{Preliminaries}
\label{Preliminaries}
The present section collects some basic facts about PSWFs and the maximum correntropy criterion.
We first introduce some mathematical notations throughout the paper.
Vectors will be denoted as the boldface lowercase letters, i.e., $\mathbf{x}$.
Matrices will be denoted by the boldface uppercase letters, i.e., $\mathbf{A}$.
The $i$-th component of $\mathbf{x}$ is $x_i$ and the $i$, $j$ element of $\mathbf{A}$ is $(\mathbf{A})_{ij}$.

%-------------------------------------------------------------------------
\subsection{Prolate Spherical Wave Functions}
\label{PSWFs}
Finding the most energy concentrated signals both in fixed time and frequency domains at the same time
is a fundament problem in information theory \cite{SP1961, MC2004, S1983}.
The problem was studied by Slepian et al. in the early's 1960 \cite{SP1961, LP1961, LP1962}, and the solutions are  the prolate spherical wave functions (PSWFs). In this part, we review the basic facts about this functions in continuous and discrete cases.

\subsubsection{Continuous Case: }

The continuous PSWFs $\{\varphi_n\}_{n=0}^\infty$ are solutions of the integral equation
\begin{eqnarray}\label{EqPSWFs}
\int_{-\tau}^{\tau}\varphi_n(s)\frac{\sin \sigma(t-s)}{\pi (t-s)}ds=\alpha_n \varphi_n(t),
\end{eqnarray}
where $[-\tau, \tau]$ and $[-\sigma, \sigma]$ are the fixed time and frequency domains, respectively.
The continuous PSWFs have several interesting properties, which follow form the general theory of integral equations and the work by Slepian \emph{et al}.
\cite{JJ2011,P1977,K1992,M2013}. We list some of them here.

\begin{itemize}
\item {\bf Eigenvalue:} The equation (\ref{EqPSWFs}) has solutions only for certain real values $\alpha_n$ of $\alpha$, and can be ordered as
$$1>\alpha_0>\alpha_1>v_2>\cdot\cdot\cdot\rightarrow 0,~~n\rightarrow\infty. $$

\item {\bf Double orthogonality:} To each $\alpha_n$ there corresponds only one eigenfunction $\varphi_n$.
The functions $\{\varphi_n\}_{n=0}^{\infty}$ form dual real orthogonal set both in the interval $(-\infty, \infty) $ and $(-\tau,\tau)$,
\begin{eqnarray}
\int_{-\tau}^{\tau}\varphi_m(t)\varphi_n(t)dt&=&\alpha_n \delta_{mn},\\
\int_{-\infty}^{\infty}\varphi_m(t)\varphi_n(t)dt&=&\delta_{mn}.
\end{eqnarray}
Here, $\delta_{mn} $ is the Delta function, i.e., $\delta_{mn}=0$ if $m \neq n$ and $\delta_{mn}=1$ for $m=n$.

\item {\bf Completeness:} A bandlimited function $y$ with its Fourier transform support on $[-\sigma,\sigma]$, can be expressed as
\begin{eqnarray}
y(t)=\sum_{n=0}^{\infty}a_n\varphi_n(t),
\end{eqnarray}
where $a_n:=\int_{-\infty}^{\infty}y(t)\varphi_n(t)dt$.

\item {\bf Fourier transform pair \cite{P1977}:} The PSWFs $\varphi_n$ and their Fourier transforms have the following relationships
\begin{eqnarray}
\varphi_n(t)&\leftrightarrow&(-i)^n\sqrt{\frac{2\pi\tau}{\sigma \lambda_n}}\varphi_n \left(\frac{\tau}{\sigma}\omega \right)p_{\sigma}(\omega),\\
\varphi_n(t)p_{\tau}(t)&\leftrightarrow&(-i)^n\sqrt{\frac{2\pi\tau}{\sigma }}\varphi_n \left(\frac{\tau}{\sigma}\omega \right)p_{\sigma}(\omega),
\end{eqnarray}
where $p_{\tau}$ is a characteristic function on $(-\tau, \tau)$,
 i.e., $p_{\tau}(t)=1$ for $t\in (-\tau, \tau)$ and $p_{\tau}(t)=0$ for $t\notin (-\tau, \tau)$.
 $p_{\sigma}$ is a characteristic function on $(-\sigma, \sigma)$,
 i.e., $p_{\sigma}(\omega)=1$ for $\omega\in (-\sigma, \sigma)$ and $p_{\sigma}(\omega)=0$ for $\omega\notin (-\sigma, \sigma)$.
\end{itemize}

%%%%%%%%%%%%%%%%%%%%%%%%%%%%
\subsubsection{Discrete Case: }

For a discrete $2M+1$ prolate spheroidal wave sequence $\{\phi_n\}_{n=-M}^M$,
which is related to the following trigonometric polynomials (namely digital prolate functions) \cite{S1978, P1977}
\begin{eqnarray}
\phi(t):=\sum_{n=-M}^{M}\phi_{n}e^{in\omega_0t},
\end{eqnarray}
where $\{\phi_n\}_{n=-M}^M$ satisfied the discrete version of the integral equation Eq. (\ref{EqPSWFs})
\begin{eqnarray}\label{EqdisPSWFs}
\sum_{n=-M}^{M}\frac{\sin \omega_0\tau(n-k)}{\pi (n-k)}\phi_{k}=\lambda_k \phi_{n},~~~~~|n|\leq M,
\end{eqnarray}
where $\omega_0$ is a constant.
From the theory of linear equation, Eq. (\ref{EqdisPSWFs}) has $2M+1$ eigenvalues
$1>\lambda_0>\lambda_1>\lambda_2>\cdot\cdot\cdot>\lambda_{2M}. $
The corresponding eigenvectors $\{\phi_n^k\}$ form an orthonormal set
\begin{eqnarray}
\sum_{n=-M}^{M}\phi_n^k\phi_n^s= \left\{
\begin{array}{ll}
1 & k=s,\\
[1.5ex]
0 & k\neq s.
\end{array}\right.
\end{eqnarray}
The discrete PSWFs also have the double orthogonality
\begin{eqnarray}
\frac{1}{T}\int_{-\frac{T}{2}}^{\frac{T}{2}}\phi^k(t)\phi^s(t)dt=
\sum_{n=-M}^{M}\phi_n^k\phi_n^s
= \left\{
\begin{array}{ll}
1 & k=s,\\
[1.5ex]
0 & k\neq s.
\end{array}\right.
\end{eqnarray}
and
\begin{eqnarray}
\frac{1}{T}\int_{-\frac{\tau}{2}}^{\frac{\tau}{2}}\phi^k(t)\phi^s(t)dt=
\sum_{n=-M}^{M}\phi_n^k\phi_n^s\lambda_k
= \left\{
\begin{array}{ll}
\lambda_k & k=s,\\
[1.5ex]
0 & k\neq s.
\end{array}\right.
\end{eqnarray}

%%%%%%%%%%%%%%%%%%%%%%%%%%%%%%%%%
\subsection{Maximum Correntropy Criterion}
\label{subsec:MCC}
As a popular criterion, the mean square error (MSE) criterion has been widely used in signal processing for decades \cite{CPAW2010}.
This reason is attributed to the low complexity and the
analytical tractability of the corresponding algorithms for MSE.
For this reason, most previous signal reconstruction methods utilize MSE as the loss function.
However, since MSE only consider the second-order statistics,
it depends on the Gaussianity assumption of the noise distribution.
This makes the MSE based methods sensitive to non-Gaussian noise.
Recently, researchers developed the maximum correntropy criterion (MCC)
based on information theoretic learning (ITL),
which exhibits better robustness to non-Gaussian noise than the MSE \cite{HZH2011,EP2002,WTL2015}.

Given two scalar random variables $X$ and $Y$, the correntropy between $X$ and $Y$ is defined by \cite{AR1970}
\begin{eqnarray}
V(X,Y):=\mathbb{E}[\kappa_{\sigma}(X-Y)]=\int_{R^2} \kappa_{\sigma}(x-y)p(x,y)dxdy,
\end{eqnarray}
where $\mathbb{E}$ denotes the expectation, $p(x,y)$ denotes the joint probability density function of $X$ and $Y$ and
$\kappa_{\sigma}(x-y)$  is the Gaussian kernel function given by
\begin{eqnarray}
\kappa_{\sigma}(x-y):=\frac{1}{\sqrt{2\pi}\sigma}e^{-\frac{(x-y)^2}{2\sigma^2}}.
\end{eqnarray}
Here $\sigma$ represents the kernel scale.
In reality, the joint probability density function $p(x,y)$ is often unknown and only a finite number of samples $\{(x_i,y_i)\}_{i=1}^N$ are available.
This leads to the following sample estimator of correntropy
\begin{eqnarray}
\hat{V}(X,Y):=\frac{1}{N}\sum_{i=1}^N\kappa_{\sigma}(x_i-y_i),
\end{eqnarray}
and the correntropy induced metric (CIM) \cite{LPC2007}
\begin{eqnarray}
\text{CIM}(X,Y)
:=\left\{ \frac{1}{N}\sum_{i=1}^N\left(\kappa_{\sigma}(0)-\kappa_{\sigma}(x_i-y_i)\right)\right\}^{\frac{1}{2}}.
\end{eqnarray}
Compared to MSE, CIM can handle non-Gaussian noises and give positive performance \cite{LPC2007}.
This motivates us to utilize the CIM as data fidelity term.

%===============================
\section{Signal Reconstruction}
\label{Signal Reconstruction}
In this section, we first give a brief introduction to the subject of signal reconstruction.
Then we present the existing PSWFs based signal reconstruction methods and
propose our improved signal reconstruction methods and their corresponding algorithms.

%%%%%%%%%%%%%%%%%%%%%%%%%%%%%%%%%
\subsection{Background of Signal Reconstruction}
The problem of signal reconstruction aims to reconstruct a bandlimited signal $x(t)$
with noise $n(t)$ from some given samples of observed signal $y(t)$ \cite{P1977}.
Specifically, if $M$ samples of the observation signal $y(t)$ are taken at times $\{t_i\}_{i=1}^M$,
namely $\mathbf{y}:=(y_1,y_2,\cdots,y_M)^T\in\mathbb{R}^M$,
where $y_i:=y(t_i),~i=1,2,\cdots,M$. We would like to reconstruct the bandlimited signal $x(t)$ given by
\begin{eqnarray}
y(t)=x(t)+n(t),~~t\in \mathbb{R}.
\end{eqnarray}

The classical Shannon sampling theorem shows that the bandlimited signal $x(t)$ can be reconstructed by the samples $b_j$
\begin{eqnarray}\label{eq.xsinc}
x(t)=\sum_{n=-\infty} ^{+\infty}b_j\frac{\sin \sigma (t-t_j)}{\sigma(t-t_j)},
\end{eqnarray}
where $ \sigma$ is a constant related to bandwidth.
In reality only finite number of samples are available, therefore we consider the finite sum related to sinc-functions $\hat{x_s}(t)$ to approximate $x(t)$,
\begin{eqnarray}\label{eq.hatxsinc}
\hat{x}_s(t) :=\sum_{j=1} ^{M}b_j\frac{\sin \sigma (t-t_j)}{\sigma(t-t_j)}.
\end{eqnarray}
Denote $\mathbf{b}:=(b_1,b_2,\cdots,b_M)^T$, $\mathbf{x}:=(\hat{x_s}_1, \hat{x_s}_2, \cdots, \hat{x_s}_M)^T \in \mathbb{R}^M$,
where $\hat{x_s}_i=\hat{x_s}(t_i),~i=1,2,\cdots,M$ and
 \begin{eqnarray}
(\mathbf{A})_{ij} := \frac{\sin \sigma (t_i-t_j)}{\sigma(t_i-t_j)},~~~~i,~j=1,2,...,M,
\end{eqnarray} Eq. (\ref{eq.hatxsinc}) can be written in matrix form as $\mathbf{Ab}=\mathbf{x}$.
If $\mathbf{b}$ is given, following Eq. (\ref{eq.hatxsinc}), then $\hat{x_s}(t)$ is the linear combination of sinc-functions.
As an approximation to $x(t)$, the mean-square error (error) between $\hat{x_s}(t)$ and $x(t)$ is given by
\begin{eqnarray*}
error:=\int_{-\infty}^{+\infty}|x(t)-\hat{x_s}(t)|^2dt= C\sum_{-\infty<j<1,j>M}|b_j|^2,
\end{eqnarray*}
where $C$ is a constant.

Giving the observed points $\mathbf{y}=(y_1,y_2,\cdots,y_M)^T\in\mathbb{R}^M$ of $y(t)$, they are the vectors $\mathbf{x}$ combining with the white noise, the method of linear least squares is a standard
approach to minimize the residual
\begin{eqnarray}\label{eq.MSE1}
\mathbf{b} =\argmin_{ \mathbf{b} \in \mathbb{R}^M} ||\mathbf{Ab}-\mathbf{y}||_2^2,
\end{eqnarray}
where $ ||\cdot||_2$ is the $\ell_2$ norm.
The solution of this problem (\ref{eq.MSE1}) is \cite{R1970, H1998}
\begin{eqnarray}\label{eq.solutionb}
\mathbf{b}=(\mathbf{A}^{T}\mathbf{A})^{-1}\mathbf{A}^{T}\mathbf{y}.
\end{eqnarray}
Utilizing Eq. (\ref{eq.solutionb}),  a linear formula
$\hat{x_s}(t)=\sum_{j=1} ^{M}b_j\frac{\sin \sigma (t-t_j)}{\sigma(t-t_j)}$ is obtained,
where $b_j$ is the $i$-th component of $\mathbf{b}$.

\subsection{Reconstruction Using PSWFs}
%Since we have introduced the special functions PSWFs and the fact in \cite{WS2003,MZ2014,SCD2009,DK2016}
%for the relationship of sinc-functions and PSWFs,
%which leads to the signal $x(t)$ reconstructed by PSWFs.
The idea for $x(t)$ reconstructed by PSWFs was used in \cite{SCD2009} already.
However, they just get the reconstruction algorithm using
the mean square error (MSE) criterion as the loss function,
which will be introduced in the subsection \ref{sec.Tikhonov Regularization} in detail.
Since the PSWFs are also used in their method, for completeness of the presentation,
we list some basic facts for the PSWFs in reconstruction problem.

Using the relationship between sinc-functions and PSWFs \cite{DK2016}
$
\frac{\sin \sigma (t-t_j)}{\sigma(t-t_j)}=\sum_{m=-\infty} ^{+\infty}\phi_m{(t)}\phi_m{(t_j)},
$
Eq. (\ref{eq.xsinc}) can be expressed by
$
x(t)=\sum_{j=-\infty} ^{+\infty}b_j\sum_{i=-\infty} ^{+\infty}\phi_i{(t)}\phi_i{(t_j)}
=\sum_{j=-\infty} ^{+\infty}\left(\sum_{i=-\infty} ^{+\infty}b_j\phi_i{(t_j)}\right)\phi_i{(t)}
=\sum_{j=-\infty} ^{+\infty}c_j\phi_j{(t)},
$
where $c_j:=\sum_{i=-\infty} ^{+\infty}b_j\phi_i{(t_j)}$. Consider the finite sum of the above series,
\begin{eqnarray}\label{eq.hatxpswfs}
\hat{x_{\phi}}(t):=\sum_{j=1} ^{N}c_j\phi_j{(t)}.
\end{eqnarray}
If the coefficients $c_j$, $j=1,2,\cdots,N$ of the linear system (\ref{eq.hatxpswfs}) are known,
 then $\hat{x_{\phi}}(t)$ can be represented as the linear combination of PSWFs.
 To find the coefficients $c_j$, $j=1,2,\cdots,N$, we first denote the coefficients vector $\mathbf{c}:=(c_1,c_2,\cdots,c_N)^T$,
 and solve the following problem
\begin{eqnarray}
\mathbf{c} =\argmin_{ \mathbf{c} \in \mathbb{R}^N} ||\mathbf{Dc}-\mathbf{y}||_2^2,
\end{eqnarray}
where $\mathbf{D}\in \mathbb{R}^{M\times N}$ and
\begin{eqnarray}
(\mathbf{D})_{ij}:=\phi_j(t_i),~~~~i=1,2,...,M,~j=1,2,...,N.
\end{eqnarray}
The solution of this problem is
\begin{eqnarray}\label{eq.solutionc}
\mathbf{c}=(\mathbf{D}^{T}\mathbf{D})^{-1}\mathbf{D}^{T}\mathbf{y}.
\end{eqnarray}
Therefore we obtain a linear formula for $\hat{x_{\phi}}$,
$\hat{x_{\phi}}(t)=\sum_{j=1} ^{M}c_j\phi_j{(t)}, $
where $c_j$ is the $i$-th component of $\mathbf{c}$.

Notice that the number of term in Eq. (\ref{eq.hatxsinc}) and (\ref{eq.hatxpswfs}) are $M$ and $N$, respectively.
The difference cames from the number of the sampling points.
The number of samples is $M$, which means the number of different time also $M$, i.e.,  $\{t_i\}_{i=1}^M$.
This leads to the Eq. (\ref{eq.hatxsinc}) has $M$ terms,
while Eq. (\ref{eq.hatxpswfs}) can choose different $N$ terms, i.e., the number of PSWFs used can be determined by ourself. Due to the energy concentration property of PSWFs, the number $N$ can be choosn small such that
the method preserves most of the energy of the signal.

%==============================================================
\subsection{Reconstruction Under Regularization and PSWFs}
\label{sec.Tikhonov Regularization}
In most of the time the solutions in Eq. (\ref{eq.solutionb}) and Eq. (\ref{eq.solutionc}) may not exist in reality,
because the inverse $(\mathbf{A}^{T}\mathbf{A})^{-1}$ and $(\mathbf{D}^{T}\mathbf{D})^{-1}$ may not exist.
This leads to the ill-posed problems \cite{CPAW2010, R1970, H1998}.
In these cases, regularization methods are needed to obtain the meaningful solutions.
In the following, we will introduce the {\bf Tikhonov regularization based reconstruction algorithm}
to overcome the mentioned ill-posed problems.

In \cite{SOC2012}, Senay \emph{et al}. proposed a method based on the Tikhonov regularization \cite{OKCN2000, T1963}
 and used the mean square error (MSE) criterion as the cost functions
to approximate the target vector $\mathbf{c}$, i.e.,
\begin{eqnarray}\label{eq.MSE2}
\mathbf{c}=\argmin_{\mathbf{c} \in \mathbb{R}^N} ||\mathbf{Dc}-\mathbf{y}||_2^2+\lambda ||\mathbf{c}||_2^2,
\end{eqnarray}
where $ ||\cdot||_2$ is the $\ell_2$ norm and $\lambda$ is the regularization parameter.
The explicit solution for this problem (\ref{eq.MSE2}) is
\begin{eqnarray}
\mathbf{c}=(\mathbf{D}^{T}\mathbf{D}+\lambda \mathbf{I})^{-1}\mathbf{D}^{T}\mathbf{y},
\end{eqnarray}
where $\mathbf{I}$ is the identity matrix.
Senay \emph{et al}. \cite{SOC2012} used the PSWFs $\mathbf{D}$ to obtain the Tikhonov regularization reconstruction (namely, RPSWF).

In the present paper, we compare our method with their RPSWF method.
%and denote the \emph{RPSWF} as the regularized PSWFs method.
Of course, if the sinc-functions $\mathbf{D}$ is applied to get the Tikhonov regularization reconstruction, we name it RSinc.

\subsection{Reconstruction Under Entropy and PSWFs}

The mean square error (MSE) criterion in the Eq. (\ref{eq.MSE2}) is known to rely to the problem with Gaussian noise assumption \cite{ LPC2007, HZH2011}.
The vast amount of noise doesn't satisfy this assumption, which leads to the poor reconstruction performance.
To overcome this problem, a maximal correntropy based reconstruction method is proposed in this paper.
We also noted them as the entropy based methods, such as entropy based PSWFs (EPSWF) and entropy based sinc-functions (ESinc).

We present the signal reconstruction method for entropy based PSWFs (EPSWF) in the following.

Suppose $M$ samples of the observation signal $y(t)$ are taken at times $\{t_i\}_{i=1}^M$.
Denote by $\mathbf{y}=(y_1,y_2,\cdots,y_M)^T\in\mathbb{R}^M$. Firstly, we use the PSWFs $\{\phi_j(t)\}_{j=1}^N$ to construct the dictionary matrix
by defining $(\mathbf{D})_{ij}=\phi_j(t_i)$ for $i=1,2,\cdots,M$ and $j=1,2,\cdots,N$.
For ease of presentation, denote by $\mathbf{d}_i$ the $i$-th row of $\mathbf{D}$.

Secondly, the coefficient vector $\mathbf{c}$ is computed by minimizing
\begin{equation}\label{eq:objective}
\mathbf{c}=\argmin_{\mathbf{c}\in \mathbb{R}^N}
\sum_{i=1}^M \left(1-\kappa_{\sigma}\left(y_i-\mathbf{d}_i\mathbf{c}\right)\right)+\lambda\|\mathbf{c}\|_2^2.
\end{equation}

\noindent\rule{1.0\textwidth}{1.5pt}

\noindent \textbf{Algorithm 1}  Signal reconstruction via EPSWF

\vspace{-0.1in}

\noindent\rule{1.0\textwidth}{1pt}

\noindent {\bfseries Input:} The vector $\mathbf{y}\in\mathbb{R}^M$ of samples with
$y_i=y(t_i)$, $i=1,2,\cdot\cdot\cdot,M$,  and the regularization parameter $\lambda$.

\noindent{\bfseries Output:} The recovered signal $\hat{x_{\phi}}(t)$.

\begin{itemize}
\item[1:] Construct the matrix  $\mathbf{D}\in\mathbb{R}^{M\times N}$
          by defining $\mathbf{D}_{ij}=\phi_j(t_i)$ for $i=1,2,\cdots,M$ and $j=1,2,\cdots,N$.

          Denote by $\mathbf{d}_i$ the $i$-th row of $\mathbf{D}$.

\item[2:] Compute the reconstruction coefficient by solving the following optimization problem
\begin{equation}
\mathbf{c}=\argmin_{\mathbf{c}\in \mathbb{R}^N}
\sum_{i=1}^M \left(1-\kappa_{\sigma}\left(y_i-\mathbf{d}_i\mathbf{c}\right)\right)+\lambda\|\mathbf{c}\|_2^2
\end{equation}

\item[3:] Calculate the recovered signal $\hat{x_{\phi}}(t)=\sum_{j=1}^N c_j\phi_j(t)$.

\end{itemize}
\noindent\rule{1.0\textwidth}{1pt}

After obtaining the coefficient vector $\mathbf{c}$, the reconstructed signal
is given by
\begin{eqnarray}
\hat{x_{\phi}}(t)=\sum_{j=1}^N c_j\phi_j(t).
\end{eqnarray}
The complete reconstruction algorithm is summarized in Algorithm 1.

We utilize the half-quadratic theory \cite{NN2005} to design the optimization strategy
to solve the problem in Eq. (\ref{eq:objective}).
According to the convex optimization theory \cite{WTL2015,R1970},
there exists a convex function $\alpha(u),~u\in\mathbb{R}$ such that
\begin{eqnarray}
\kappa_{\sigma}(t)=\sup\left\{\frac{ut^2}{\sigma^2}-\alpha(u),~u\in \mathbb{R}_-\right\},
\end{eqnarray}
where $u=-\kappa_{\sigma}(t)$ reaches the supremum. Then there holds
\begin{eqnarray}
-\kappa_{\sigma}(t)=\inf\left\{-\frac{ut^2}{\sigma^2}+\alpha(u),~u\in \mathbb{R}_-\right\}.
\end{eqnarray}
If we define $w=-\frac{u}{\sigma^2}$ and $\beta(w)=\alpha(u)$, we have
\begin{eqnarray}
-\kappa_{\sigma}(t)=\inf\left\{wt^2 +\beta(w),~w\in \mathbb{R}_+\right\},
\end{eqnarray}
where the infimum is reached at $w=\frac{1}{\sigma^2}\kappa_{\sigma}(t)$.

Applying the property  in Eq.(\ref{eq:objective}) and removing constants,
we can reformulate the problem in Eq. (\ref{eq:objective}) as
\begin{equation}\label{eq:CPSWF2}
\min_{\mathbf{c}\in\mathbb{R}^N, \mathbf{w}\in\mathbb{R}^M_+} J(\mathbf{c}, \mathbf{w})=
 \sum_{i=1}^M \left(w_i\left(y_i-\mathbf{d}_i\mathbf{c}\right)^2
+\beta(w_i)\right)+\gamma \|\mathbf{c}\|_2^2,
\end{equation}
where $\mathbf{w}=(w_1, w_2, \cdots,  w_M)^T \in\mathbb{R}^M$ is a vector composed of auxiliary variables.
A local minimizer of problem in Eq. (\ref{eq:CPSWF2}) can be obtained by alternatively updating $\mathbf{c}$ and
$\mathbf{w}$. Specifically, while fixing the coefficient vector $\mathbf{c}$,  the auxiliary vector $\mathbf{w}$
can be updated by setting $w_i^{(t+1)}=\frac{1}{\sigma^2}\kappa_{\sigma}\left(y_i-\mathbf{d}_i\mathbf{c}^{(t)}\right),
~i=1, 2, \cdots, M$ according to the analysis above.
Here $t$ is the number of iterations.
While fixing $\mathbf{w}$, the problem in Eq. (\ref{eq:CPSWF2}) is equivalent to
\begin{eqnarray}\label{eq:updatec}
\mathbf{c}^{(t+1)} = \argmin_{\mathbf{c}\in \mathbb{R}^n}\left\|\sqrt{\text{diag}\left(\mathbf{w}^{(t+1)}\right)}\mathbf{y}-
\sqrt{\text{diag}\left(\mathbf{w}^{(t+1)}\right)}\mathbf{D}\mathbf{c}\right\|_2^2+\lambda\|\mathbf{x}\|_2^2.
\end{eqnarray}
where $\text{diag}(\mathbf{w}^{(t+1)})$ denotes a square diagonal matrix
with the elements of $\mathbf{w}^{(t+1)}$ on the main diagonal.

The optimization problem in Eq. (\ref{eq:updatec}) has a close form solution, which can be
explicitly expressed as
\begin{eqnarray}
\mathbf{c}^{(t+1)}=\left(\mathbf{D}^T\text{diag}\left(\mathbf{w}^{(t+1)}\right)\mathbf{D}
+\lambda \mathbf{I}\right)^{-1}\mathbf{D}^T\text{diag}\left(\mathbf{w}^{(t+1)}\right)\mathbf{y}.
\end{eqnarray}
As for the kernel size $\sigma$, it is determined empirically \cite{P2010} and set as
\begin{eqnarray}
\sigma = \left(\frac{1}{2M}\left\|\mathbf{y}-\mathbf{D}\mathbf{c}\right\|_2^2\right)^{\frac{1}{2}}.
\end{eqnarray}
Algorithm 2 summarizes the complete procedure for solving the problem  in Eq. (\ref{eq:objective}).
In light of the half-quadratic theory \cite{NN2005},
the sequence $\left\{J(\mathbf{c}^{(t)}, \mathbf{w}^{(t)})\right\}_{t=1}^{\infty}$ always converges.

\noindent\rule{1.0\textwidth}{1.5pt}

\noindent \textbf{Algorithm 2}  Solving the optimization problem  in Eq. (\ref{eq:objective})

\vspace{-0.1in}
\noindent\rule{1.0\textwidth}{1pt}

\noindent{\bfseries Input:} $\mathbf{y}$, $\mathbf{D}$, $\lambda$.

\noindent{\bfseries Output:} $\mathbf{c}$.

\noindent {\bfseries Repeat until convergence:}
\begin{itemize}
\item[1:] Update the auxiliary variables $\{w_i\}_{i=1}^{i=M}$
\begin{eqnarray}
w_i^{(t+1)}=\frac{1}{\sigma^2}\kappa_{\sigma}\left(y_i-\mathbf{d}_i\mathbf{c}^{(t)}\right),
  ~i=1, 2, \cdots, M.
\end{eqnarray}

\item[2:] Update the coefficient vector $\mathbf{c}$
\begin{equation}\label{eq:updatec}
\mathbf{c}^{(t+1)} = \argmin_{\mathbf{c}\in \mathbb{R}^N}\left\|\sqrt{\text{diag}(\mathbf{w}^{(t+1)})}\mathbf{y}-
\sqrt{\text{diag}(\mathbf{w}^{(t+1)})}\mathbf{D}\mathbf{c}\right\|_2^2+\lambda\|\mathbf{x}\|_2^2.
\end{equation}
\end{itemize}
\noindent\rule{1.0\textwidth}{1pt}

\begin{figure}[!t]
  \centering
 \includegraphics[height=6.2cm]{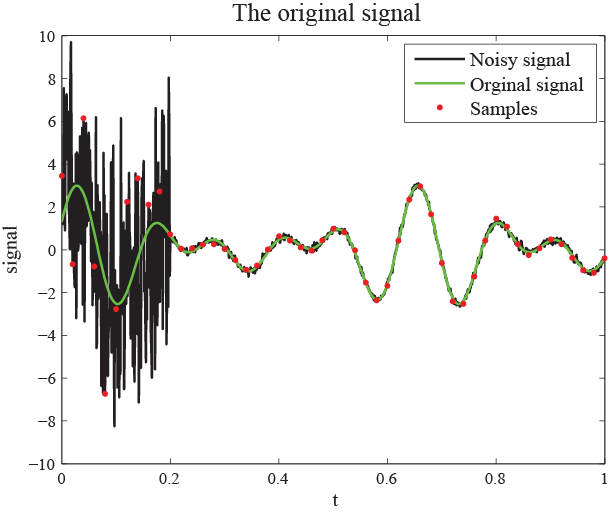}
  \par
 \includegraphics[height=6.2cm]{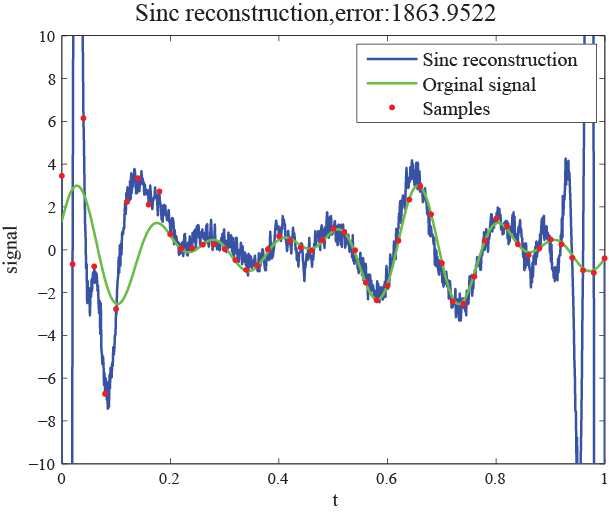}
  \includegraphics[height=6.2cm]{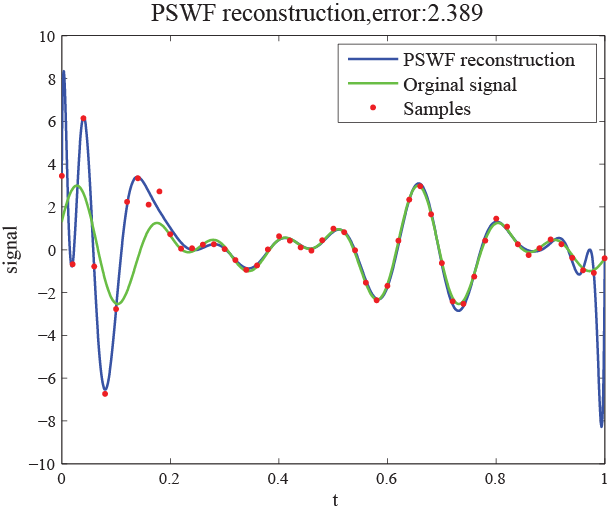}
 \par
 \includegraphics[height=6.2cm]{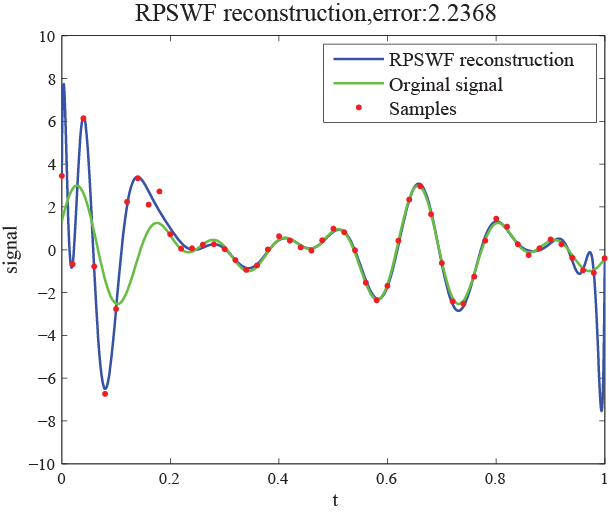}
  \includegraphics[height=6.2cm]{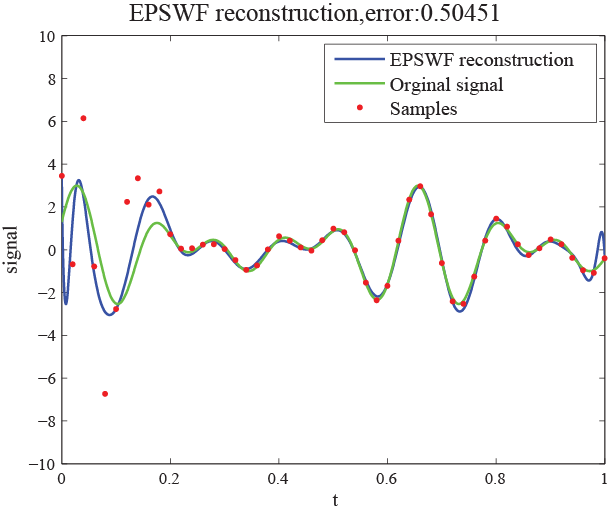}
  \caption{ Results for uniformly sampled signal reconstruction and error.
  The first image shows the original signal with noise and the samples.
  The following $4$ images show the reconstructed signals in blue lines with different methods.
  The Sinc-functions method with reconstructed error $1863.9522$;
  the PSWF method with reconstructed error $2.3890$;
  the RPSWF method with reconstructed error $2.2368$;
  the EPSWF method with reconstructed error $0.5045$.}
  \label{Figureuniformlyreconstruction}
\end{figure}

\section{Experiments}
\label{Experiments}

\begin{figure}[!t]
  \centering
 \includegraphics[height=6.2cm]{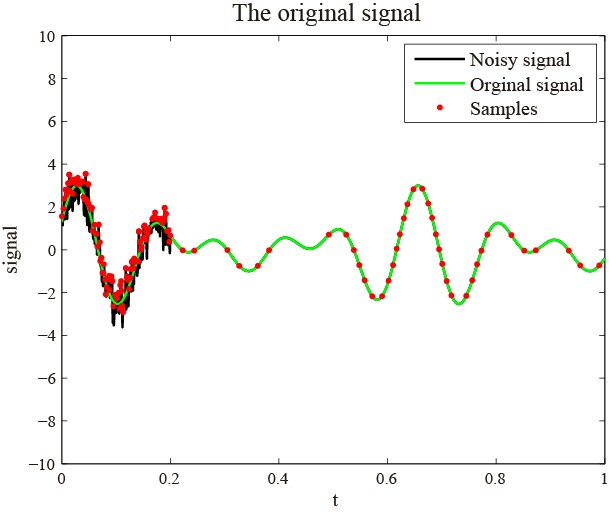}
  \par
 \includegraphics[height=6.2cm]{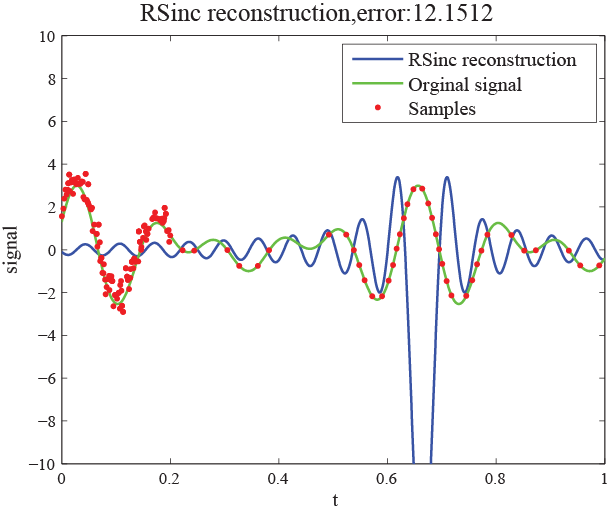}
  \includegraphics[height=6.2cm]{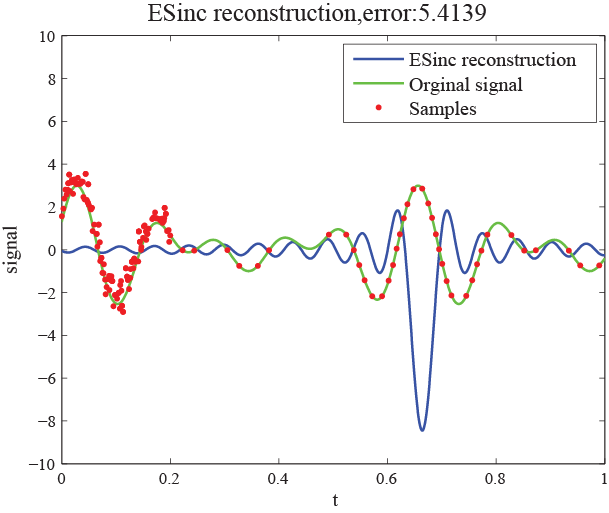}
 \par
 \includegraphics[height=6.2cm]{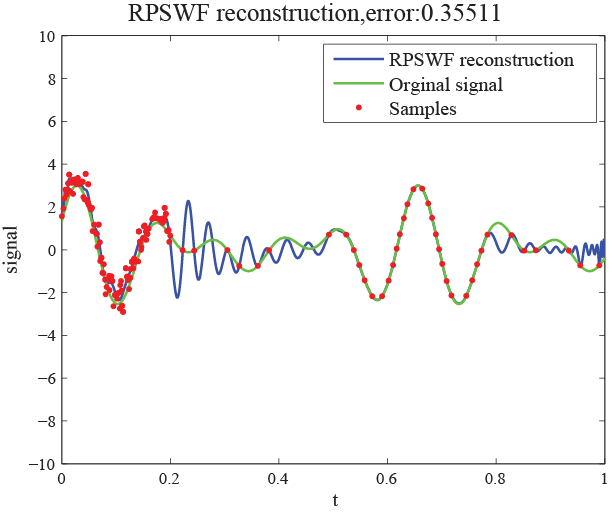}
  \includegraphics[height=6.2cm]{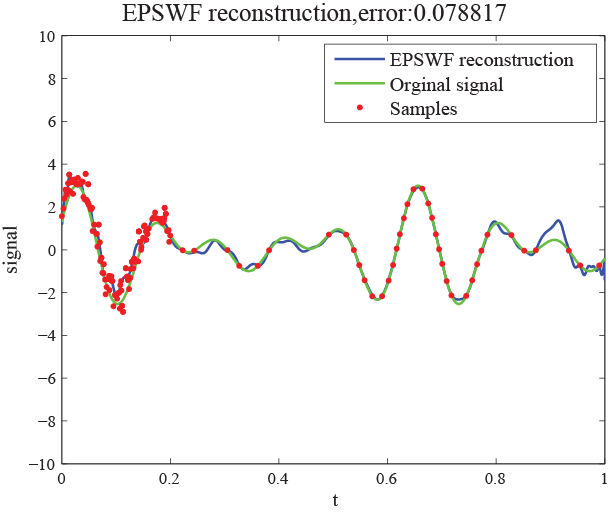}
  \caption{ Results for nonuniformly sampled signal reconstruction and error.
  The first image shows the original signal with noise and the samples.
  The following $4$ images show the reconstructed signals in blue lines with different methods.
  The RSinc method with reconstructed error $12.1515$;
  the ESinc method with reconstructed error $5.4139$;
  the RPSWF method with reconstructed error $0.3551$;
  the EPSWF method with reconstructed error $0.0788$.
  }
  \label{Figurenonuniformlyreconstruction}
\end{figure}

In this section, we present the the performance of the proposed methods by applying them to a  signal $x(t)$.
The signal $x(t)$ is a combination of three sinusoids original signals  embedded in noise, i.e.,
\begin{eqnarray}
x(t)=sin(50t+0.1)+sin(30t+0.8)+sin(40t+0.5),~~0\leq t \leq 1.
\end{eqnarray}
In Fig. \ref{Figureuniformlyreconstruction} and \ref{Figurenonuniformlyreconstruction},
the green line is the original signal $x(t)$ for $0\leq t \leq 1$.

The experiments include two parts, the first one is about the signal with a large quantity of noise in  $0\leq t \leq 0.2$,
which is shown in Fig. \ref{Figureuniformlyreconstruction} the black line.
While the second experiment add a small quantity of noise in  $0\leq t \leq 0.2$,
which is shown in Fig. \ref{Figurenonuniformlyreconstruction} the black line.

In the first experiment, we uniformly sample some sample points, which is shown in red points in Fig. \ref{Figureuniformlyreconstruction}.
Since the large noise added in the signal, some of samples in $0\leq t \leq 0.2$ are far away from the original signal.
The blue lines in  Fig. \ref{Figureuniformlyreconstruction} shows the reconstructed results for different methods.
In more specific terms, the methods include Sinc, PSWF, RPSWF and EPSWF. The reconstructed error are also shown in these pictures.
From the Fig. \ref{Figureuniformlyreconstruction}, we can obtain the following conclusions:
\begin{itemize}
  \item The PSWFs based method is greater than the sinc-functions based method.
  \item The Tikhonov regularization based reconstruction method (RPSWF) is better than the non-regularization methods (PSWF and sinc-functions).
  \item The maximal correntropy based Reconstruction method (EPSWF) is the best method among all of the methods.
\end{itemize}
 However, we can find the the reconstruction error for RPSWF $2.2368$  is not much to improve than that of  PSWF method $2.3890$.
 While, the reconstruction error for EPSWF $0.5045$ is much smaller than $2.2368$.
 This results verify the superiority of EPSWF for signal with large noise.

In the second experiment, we non-uniformly sample some sample points,
which is shown in red points in Fig. \ref{Figurenonuniformlyreconstruction}.
The samples in $0\leq t \leq 0.2$ are intensive and the samples in $0.2\leq t \leq 1$ are sparse.
The blue lines in  Fig. \ref{Figurenonuniformlyreconstruction} shows the reconstructed results for different methods.
Since we have known that the regularization based and maximal correntropy based reconstruction methods have good performance,
 we only compare these methods in this experiments.
In more specific terms, the methods include RSinc, PSinc, RPSWF and EPSWF. The reconstructed error are shown in pictures.
From the Fig. \ref{Figurenonuniformlyreconstruction}, we can obtain the following conclusions:
\begin{itemize}
  \item The RPSWF based method is significantly better than the RSinc based method
  and the EPSWF based method is far better than the ESinc based method,
  i.e., the PSWFs based method is better than sinc-functions based method.
  \item The EPSWF based method is far better than the RPSWF based method and
  ESinc based method is far better than the RSinc based method,
  i.e., the maximal correntropy based reconstruction methods is better than the regularization based reconstruction methods.
\end{itemize}

\section{Conclusions}
\label{sec.Conclusions}

In this paper, we proposed a novel robust signal reconstruction method based on the
prolate spherical wave functions (PSWFs) and maximum correntropy criterion (MCC).
The PSWFs have been proven to have good performance in signals  representation.
But the existed PSWFs method only consider the MSE criterion method,
which has good performance for noise obey the Gaussian distribution.
For the impulsive noise and outliers, the MSE based method leads to large reconstruction error.
For these reasons, we proposed the MCC based PSWFs reconstruction method (EPSWF).
The experimental results on synthetic signals show that the EPSWF can obviously improve
the performance in signal reconstruction.
%%%%%%%%%%%%%%%%%%%%%%

\end{document}